%% file: main.tex
\documentclass[aps,prb,amsmath,amssymb,twocolumn,floatfix,showpacs,superscriptaddress,bibliography]{revtex4-1}

\usepackage{graphicx}
\usepackage[usenames,dvipsnames,svgnames,table]{xcolor}
\usepackage{hyperref}
\hypersetup{
	colorlinks=true,       
	linkcolor=VioletRed,  
    citecolor=JungleGreen,        
    filecolor=Orchid,      
    urlcolor=black           
}

\usepackage{hhline} 
\usepackage{epstopdf}
\usepackage{times}
\usepackage{multirow}
\usepackage[utf8]{inputenc}

\usepackage{tikz}
\usepackage{pgfplots}
\pgfplotsset{compat=1.14}

\newcommand{\Renyi}{R\'{e}nyi\ }
\DeclareMathOperator{\Tr}{Tr}

\begin{document}

\title{Universal divergence of the \Renyi entropy of a thinly sliced torus at the Ising fixed point}

\author{Bohdan Kulchytskyy}
\affiliation{Perimeter Institute for Theoretical Physics, Waterloo, Ontario, N2L 2Y5, Canada}
\affiliation{Department of Physics and Astronomy, University of Waterloo, Ontario, N2L 3G1, Canada}

\author{Lauren E. Hayward~Sierens}
\affiliation{Perimeter Institute for Theoretical Physics, Waterloo, Ontario, N2L 2Y5, Canada}

\author{Roger G. Melko}
\affiliation{Perimeter Institute for Theoretical Physics, Waterloo, Ontario, N2L 2Y5, Canada}
\affiliation{Department of Physics and Astronomy, University of Waterloo, Ontario, N2L 3G1, Canada}

\date{\today}

\begin{abstract}

\input{Abstract}

\end{abstract}

\maketitle

\input{Introduction}

\input{Theory}

\input{Methods}

\input{Results}

\input{Discussion}

\section*{Acknowledgments}

We acknowledge crucial discussions with W. Witczak-Krempa, R. Myers,  C. Herdman, A. Schlief and M. Metlitski on these and related ideas.
We thank the Galileo Galilei Institute for Theoretical Physics for hospitality, and the INFN for partial support during the completion of this work.
Computer simulations were made possible by the facilities of the Shared Hierarchical Academic Research Computing Network (SHARCNET) and Compute/Calcul Canada. 
This research was supported by the Natural Sciences and Engineering Research Council of Canada (NSERC), the Canada Research Chair program, the National Science Foundation under Grant No. NSF PHY-1748958,
and the Perimeter Institute for Theoretical Physics. Research at Perimeter Institute is supported by the Government of Canada through Industry Canada and by the Province of Ontario through the Ministry of Research \& Innovation.

\bibliographystyle{apsrev4-1}
\bibliography{References}

\end{document}

%% file: Abstract.tex
The entanglement entropy of a quantum critical system can provide new universal numbers that depend on the geometry of the entangling bipartition.
We calculate a universal number called $\kappa$, which arises when a quantum critical system is embedded on a two-dimensional torus and bipartitioned into two cylinders.
In the limit when one of the cylinders is a thin slice through the torus, $\kappa$ parameterizes a divergence that occurs in the entanglement entropy sub-leading to the area law.
Using large-scale Monte Carlo simulations of an Ising model in 2+1 dimensions, we access the second \Renyi entropy, and determine that, at the Wilson-Fisher (WF) fixed point, $\kappa_{2,\text{WF}} = 0.0174(5)$. 
This result is significantly different from its value for the Gaussian fixed point, known to be $\kappa_{2,\text{Gaussian}} \approx 0.0227998$.

%% file: Introduction.tex
\section{Introduction}

As an alternative to traditional critical exponents, universal numbers extracted from the \Renyi entanglement entropy (EE) can be used to characterize quantum critical points. 
In general, it is an open question whether the universal information about a fixed point contained in entanglement quantities differs from that contained in critical exponents.
Recently, there has been considerable progress in relating the content of universal numbers from entanglement entropies to those obtained from two-point functions in $2+1$ dimensions. A particular success story is the contribution from a local corner in the entangling boundary.  There, the universal coefficient of the corner contribution has been shown to be related to the central charge $C_T$ defined by the two-point function of the stress tensor.  This result was initiated by numerical studies of interacting quantum critical systems,\cite{Inglis_2013,Kallin:2013} where the corner contribution was observed to scale with the number of degrees of freedom of the underlying field theory. \cite{Miles:2014,Kallin:2014}  Subsequent comparison to similar scaling in $C_T$, numerically calculated from the conformal bootstrap, \cite{Kos:2014} motivated a conjecture relating the two quantities, \cite{Bueno:2015,Bueno:2015JHEP,Bueno:2016} which was eventually proven in general. \cite{Faulkner2016}
Further studies have examined the behaviour of universal corner coefficients for more general angles and \Renyi indices in $2+1$ dimensions,\cite{Helmes:2016} while other work has examined universal features due to corners in higher dimensions.\cite{Casini:2009,Kovacs2012,Klebanov:2012,Myers:2012,Devakul:2014_2,BuenoMyers2015,Hayward2017,Bednik2018}

A corner or vertex is only one geometry that induces a universal contribution to the \Renyi entanglement entropy in $2+1$ dimensions.  In this paper, we turn to a critical system defined on a torus, where the entangling region is a cylindrical slice and the \Renyi entropy contains a universal scaling coefficient that we call $\kappa$.  
Contrary to the recent efforts described above to study a corner's universal contribution to the \Renyi entropy in both free and interacting theories, studies of $\kappa$ have been relatively restricted, with results obtained only for free theories and theories with a gravitational dual. The cases where $\kappa$ has been studied include the free scalar field theory with the dynamical exponent $z=1$ \cite{Casini:2009, Bueno:2015JHEP, Hayward2016, Witczak2017} and $z=2$ (the quantum Lifshitz model), \cite{Stephan_2013, TwistTorus} as well as their fermionic analogues: Dirac fermions with $z=1$\cite{Casini:2009, Bueno:2015JHEP, Chen_2015} and $z=2$ (the quadratic band touching model).\cite{Chen_2015, TwistTorus}

For CFTs in $2+1$, the universal coefficient $\kappa$ is known to be rigorously related to the universal corner coefficient through an exact conformal mapping in the thin-slice and small-angle limit of the cylindrical and corner geometries, respectively.\cite{Casini:2009, CornerCylinder} In general, the similarity between these two universal coefficients seems to extend well beyond this limit (and to other models).\cite{Witczak2017} However, while cylindrical entanglement geometries are amenable to lattice calculations, corner contributions to the entanglement entropy are notoriously difficult to measure in lattice calculations beyond certain special angles that are natural to the lattice.\cite{Helmes:2016} 

 In $(2+1)$-dimensional interacting theories, no exact results for $\kappa$ have been obtained to date.  Here, we examine the cylindrical-slice \Renyi EE in a $(2+1)$-dimensional Ising model, an interacting theory for which the critical behaviour is governed by the scalar Wilson-Fisher fixed point. 
To extract universal characterizations of the Wilson-Fisher fixed point from the EE, we utilize a mapping from the quantum theory in $2+1$ dimensions to a 3-dimensional critical Ising model. 
Such quantum to classical mappings are well-understood for the purposes of extracting universal critical exponents, but are much less  explored for extracting universal numbers from EE.\cite{Caraglio2008,typicality,Alba2010}
This mapping allows us to utilize Monte Carlo simulations of a critical isotropic $3$-dimensional classical Ising model, which can be tuned to a thermal phase transition described by the Wilson-Fisher fixed point.
We numerically access the second \Renyi EE, 
for which the scaling has been shown to behave qualitatively similarly to that of the more familiar (von Neumann) EE,\cite{Miles:2014,Kallin:2014} 
by confining our Ising model to a two-sheeted Riemann surface \cite{Melko:2010,Calabrese:2004}.
We devise an improved estimator \cite{Caraglio2008, Bohdan} based on the Fortuin-Kasteleyn cluster representation of the Ising model, \cite{FK, random_cluster}
which allows us to obtain a high-quality dataset.
While these simulations allow us to access the \Renyi EE numerically, strong lattice effects occur in the entanglement scaling, which can significantly bias extrapolated estimates if not accounted for properly. 
We thus develop a novel fitting procedure that makes it possible to accurately probe the thin-slice scaling limit of the EE on the lattice.
The procedures we introduce are crucial for obtaining our final result, which reveals that while the numerical value for the universal $\kappa_2$ obtained for the Wilson-Fisher fixed point is close to the Gaussian result, it nonetheless shows a statistically significant difference.

%% file: Theory.tex
\section{Scaling Theory 
\label{Scaling}}

We consider an interacting quantum field theory regularized on a two-dimensional square lattice with lattice spacing (cutoff) $\delta$ and linear dimensions $L_x$ and $L_y$.   The theory is wrapped on a torus such that the system has periodic boundary conditions.  

\begin{figure}[t]
\includegraphics[clip, trim=0 0 0 0.5cm]{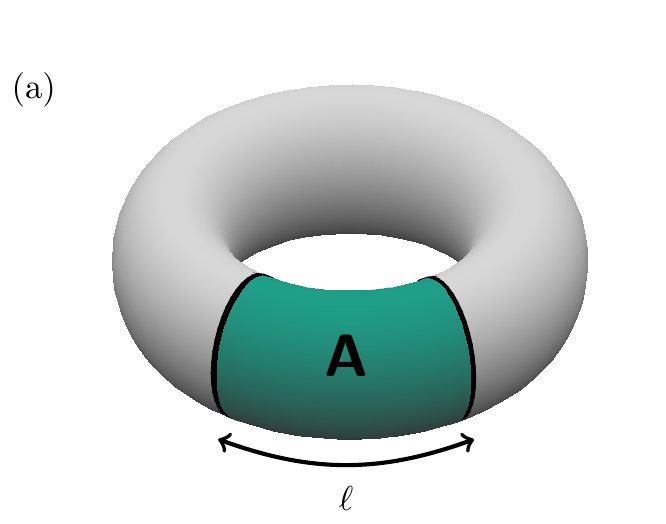} \\ 
\vspace{2mm}
\includegraphics{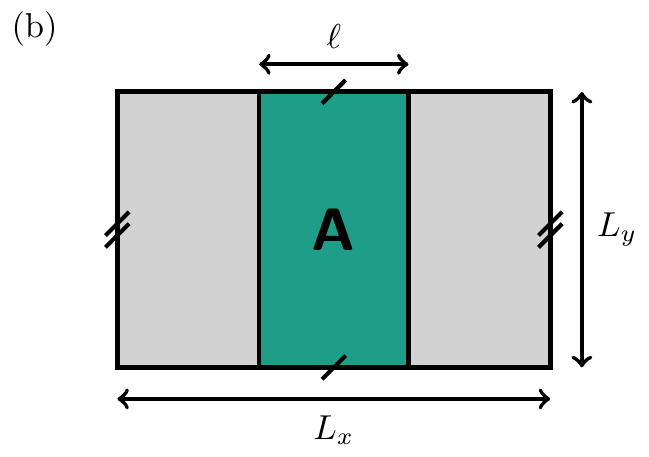}

\caption{Entanglement bipartition used to extract the universal number $\kappa$. 
The system has periodic boundary conditions such that it can be visualized on the surface of a torus as in (a), or within the two-dimensional plane as in (b).
Region $A$ is a cylindrical subregion that wraps around the $y$-direction and has length $\ell$ along the $x$-direction.
}
    \label{fig_regA_cylinder}
\end{figure}

We analyze features of the \Renyi entanglement entropy
\begin{equation}
S_n(A) = \frac{1}{1-n} \ln \big[ \Tr  \rho_A^n\big],
\label{eq:renyiEE}
\end{equation}
where $\rho_A = \Tr_{\bar A} \rho$ is the reduced density matrix for subregion $A$ with $\bar A$ being its complement, and $n$ is the \Renyi index. In the limit where $n \to 1$, Eq.~\eqref{eq:renyiEE} yields the von Neumann entropy such that $S_1(A) = - \Tr \! \big[ \rho_A \log \rho_A \big]$. Our numerical simulations focus on $S_2(A)$.
In this work, we consider cylindrical subregions $A$ of length $\ell$ as illustrated in Fig.~\ref{fig_regA_cylinder}.  

We first consider the general scaling behavior of the \Renyi entropy expected at a scale-invariant fixed point.
To do so, we employ two simple postulates, applied in the context of a renormalization group (RG) flow.
First, we assume that at each RG length scale $r$, the significant contribution to the entanglement entropy occurs local to the entangling boundary.  Second, due to scale invariance, we postulate that we must sum the EE contributions from every length scale $r$ along the renormalization group (RG) flow.  This procedure gives us a scaling form for any entangling bipartition of interest.
When applied to our cylindrical entanglement bipartition on a torus, we expect that\cite{Swingle_2010,LaurenPHD}
\begin{equation}
    \label{eq:Sn_integral}
    S_n(\ell, L_x, L_y) \propto \int_\delta^{r_\text{max}} \frac{\partial A}{r} \, \text{d}(\log r) + \ldots,
\end{equation}
where $\partial A = 2 L_y$ is the size of the entangling boundary and $r_\text{max}$ is the maximum contributing length scale along the RG flow. In general, $r_\text{max}$ depends upon the lengths $\ell$, $L_x$ and $L_y$ (and it does not depend on the lattice cutoff $\delta$). The ellipses denote further (non-universal) corrections, which are elements of $\mathcal{O}\!\left(\delta/\ell\right)$ and can be expressed as $g_n\!\left(\delta/\ell\right)~+~\ldots$. 
For our cylindrical choice of entangling region $A$, we thus expect the \Renyi entropy to obey the scaling form
\begin{equation}
    \label{eq:Sn_scaling}
	S_n(\ell, L_x, L_y) = a_n \frac{L_y}{\delta} + \chi_n \!\left(u,b \right) + g_n\!\left( \frac{\delta}{\ell} \right)+\cdots ,
\end{equation}
where $u=\ell/L_x$, $b=L_x/L_y$, and the ellipses denote further finite-size subleading corrections such as those that scale as $\delta/L$. 
The first term in this equation is the leading non-universal {\it area law},\cite{Sorkin1983,Bombelli1986,Srednicki1993} which is proportional to the boundary of region $A$.
The function $\chi_n$ is universal (cutoff-independent) and is fully determined by the geometric aspect ratios defining the system.
The so-called \textit{conical singularity} term $g_n$ arises solely for \Renyi entropies ($n\geq 2$). 
In this work, we focus on the second \Renyi entropy, for which $n=2$.

The behaviour of $\chi_n$ as a function of the aspect ration $u$ is known to obey certain restrictions. In particular, being an entropy measure of pure states, it must respect a symmetry around $u=\frac{1}{2}$ such that $\chi_n(u) = \chi_n(1-u)$. 
In the limit where $u\ll 1$ (\textit{i.e.}, $\ell \ll L_x$), the EE contributions at different length scales are expected to be indifferent to the infrared boundary conditions and we expect that we can estimate $S_n$ by considering the EE of a strip-like region embedded in infinite space.\cite{Casini:2009} In this case $r_\text{max} = \ell$ in Eq.~\eqref{eq:Sn_integral} and we expect that
\begin{equation}
S_{n}(\ell \ll L_x, L_y) = a_n \frac{L_y}{\delta} -  \frac{\kappa_{n}}{bu} + g_n\!\left( \frac{\delta}{\ell} \right)+\cdots ,
\end{equation}
where $\kappa_n$ is a universal coefficient that can vary with the \Renyi index $n$.
By comparing with Eq.~\eqref{eq:Sn_scaling}, we see that the second term in this equation reveals the behaviour of the universal scaling term for the cylinder entanglement in the thin slice limit such that 
\begin{equation}
\label{eq:Sn_thin}
\chi_n \!\left(u \to 0,b \right)
= - \frac{\kappa_{n}}{bu}.
\end{equation}

The numerical value of $\kappa_n$ has previously been calculated in a number of free theories, holographic duals, and phenomenological models. \cite{Stephan_2013, Chen_2015, TwistTorus, Hayward2016, Witczak2017, TwistTorus}  Specifically, values of $\kappa$ for the free scalar field theory in $2+1$ are $\kappa_{1,\text{Gaussian}} = 0.0397$ for the von Neumann entropy\cite{Casini:2009} and $\kappa_{2,\text{Gaussian}} = 0.0227998$ for the second \Renyi entropy $S_2(A)$. \cite{Bueno:2015JHEP}
This second value $\kappa_{2,\text{Gaussian}}$ is of particular relevance to the present study.

While the main focus of this study is to better understand the universal geometric function $\chi_n$, it is important to control the behaviour of the other terms in order to properly extrapolate to the thermodynamic limit. 
In particular, the conical singularity term $g_n(\delta/\ell)$ is unavoidable in Monte Carlo calculations of the \Renyi entropies. Physically, this conical singularity is due to the restructuring of the lattice that occurs when one calculates \Renyi entropies using the replica trick, as we will explain in Sec.~\ref{subsec:EEEstimator}.
While this modification does not change the coordination number of the lattice, it affects its topological structure, giving arise to a relevant operator that is locally confined. This effect transpires through the subleading anomaly correction term that we have called $g_n\!\left( \delta/\ell \right)$.\cite{CardyCalabrese2010}  As a recent study shows,\cite{Sahoo} when unaccounted for, the presence of this term can lead to erroneous extrapolation results. Unfortunately for us, while its scaling form is known for $1+1$ systems, no such analytical result exists in $2+1$. However, the EE contribution from this term is expected to grow when $\ell$ becomes small, which is exactly the scaling regime in which we are interested. Therefore this conical singularity term cannot be safely neglected. For this reason, we develop an extraction procedure, outlined in Sec. \ref{sec:Results}, designed to directly access the universal term $\chi_2$ by isolating the effect of the conical singularity term .

%% file: Methods.tex
\section{Monte Carlo Methods}
\subsection{Simulation Space}

In order to connect the quantum mechanical entanglement entropy to a classical computer simulation, we exploit the well-known correspondence between a $d$-dimensional quantum system and a $(d+1)$-dimensional classical path integral.  In the case of, say, the transverse-field Ising model in two spatial dimensions, one knows exactly how to map the parameters of the model to a 3-dimensional classical Ising model with anisotropic couplings in space and imaginary time. This mapping is routinely exploited in path-integral quantum Monte Carlo methods. However, one does not necessarily need to determine the exact parametric correspondence when studying critical properties. Indeed, the renormalization group guarantees that both the quantum critical point in $d$ dimensions, and the thermal phase transition in $d+1$ dimensions, are governed by the same fixed point.  Therefore, in this paper we take the strategy of directly simulating the 3-dimensional isotropic Ising model with classical spins $s \in \{-1,1\}$ and reduced Hamiltonian given by
\begin{equation}
    E(\mathbf{s}) = - J/T_c\sum_{\left< i, j\right>} s_i s_j,
\end{equation}
at its critical temperature $J/T_c = 0.2216544$ \cite{3dIsingTc}.
 
\begin{figure}[t]
   \includegraphics{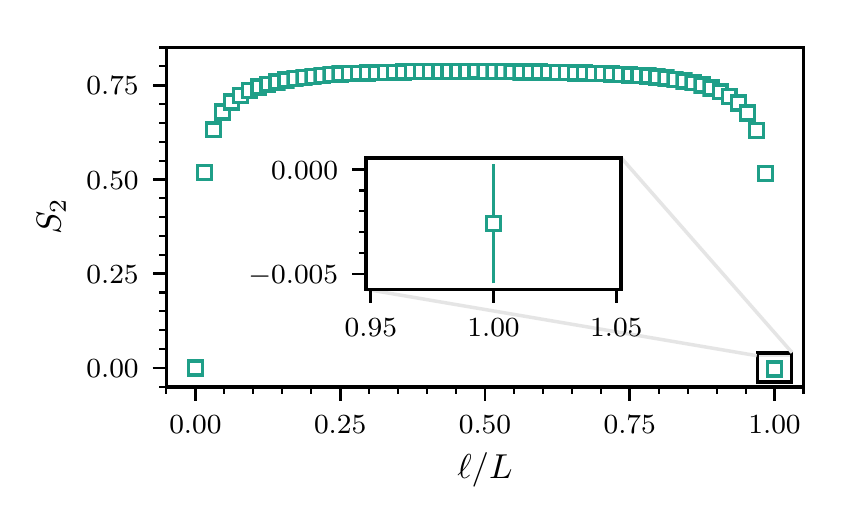}
\caption{Test of the thermal convergence of $S_2$ for a system of linear size $L=16$. The direction corresponding to the imaginary time is taken to be $L_{\tau}=15L$. The reflection symmetry of the measured $S_2$ around $\ell/L = 0.5$ is in correspondence with the theoretical expectation that the entanglement entropy of region $A$ is the same as the one of its complement in the ground state . The inset zooms in on a single datapoint corresponding to region $A$ comprising the full system. Its value is zero within the errorbar further indicating the absence of thermal fluctuations. }
    \label{ther_conv}
\end{figure}

Since the field theory of interest (the scalar $\phi^4$ theory) is Lorentz invariant with a dynamic exponent $z=1$, finite size scaling studies often scale the imaginary time dimension $L_{\tau}$ proportional to the linear spatial dimension $L$ (with a proportionality constant being close to unity for some observables~\cite{typicality}).  However, this dimension effectively represents the quantum inverse temperature, namely, $J^{Q}\beta^{Q} = L_\tau \Delta \tau$ where $\Delta \tau$ is the unitless discretization constant of the lattice in the imaginary time direction and $J^{Q}$ is the interaction parameter for the quantum model (for instance, the ferromagnetic coupling). For studies of the ground-state entanglement entropy, one must be careful to ensure that this new dimension is large enough to eliminate all thermal contributions.  For this reason, we adopt the practical strategy of converging our simulations below the energy gap $\Delta \left(L\right)$  due to the finite system size $L$, such that $\beta^Q \gg \frac{1}{\Delta \left(L\right)}$. The convergence can be tested by studying the amount of thermal entropy contained in the entire system, which should be zero if we are indeed probing the (pure) quantum ground state.  
As shown on the inset of Fig. \ref{ther_conv}, the entropy of the whole state is zero within the statistical uncertainty for a proportionality constant $L_\tau/L = 15$.   
 
\subsection{\Renyi Entropy Estimator}
\label{subsec:EEEstimator}
In order to gain quantitative access to the \Renyi EE $S_n$, we employ the replica trick, \cite{Calabrese:2004, Melko:2010} which reformulates the problem of calculating $S_n$ as finding the ratio of modified partition functions.
We introduce the notation $Z\left[n, A\right]$ to refer to a partition function defined over an $n$-sheeted Riemann space-time surface with $A$ defining the connectivity between those sheets/replicas in the imaginary time direction: spins that are part of region $A$ have inter-replica interactions while spins outside of this region have only intra-replicas interactions. One can then show that
\begin{equation}
    \label{eq:ReplicaTrick}
    S_n\left(A\right)  = \frac{1}{1-n} \log \frac{Z\left[n, A\right]}{Z\left[n, \varnothing \right]},
\end{equation}
where $Z\left[n, \varnothing \right] = Z^n$ represents $n$ independent replicas without connections between them. Two sheets are used to obtain the second \Renyi entropy $S_2$, which is the quantity of interest in our paper. 

In order to estimate the ratio of partition functions, we employ the general identity
\begin{equation}
    \label{eq:RatioTrick}
    \frac{Z'}{Z} = \left< \frac{W'(\mathbf{x})}{W (\mathbf{x})} \right>_{\mathbf{x}\sim Z},
\end{equation}
where $\mathbf{x}$ refers to a configuration, for instance a spin configuration, sampled from the distribution defined by the unnormalized probability $W(\mathbf{x})$. The partition function can be expressed as $Z=\sum_{\mathbf{x}'} W(\mathbf{x}')$, and a similar expression holds for $Z'$ in terms of $W' (\mathbf{x})$. By using the definition of the expectation value $\langle Q \rangle_{\mathbf{x}\sim W (\mathbf{x})} \equiv \frac{1}{Z}\sum_{\mathbf{x}} Q(\mathbf{x}) W(\mathbf{x})$, one can see that  Eq.~\eqref{eq:RatioTrick} holds whenever the partition functions $Z$ and $Z'$ share the same configuration space.

In our case, the configuration space is that of the Ising model with $\mathbf{x}=\mathbf{s}$ and the weights are the Boltzmann weights $e^{-\beta_c E\left(\mathbf{s}\right)}$. Thus from Eqs.~\eqref{eq:ReplicaTrick} and~\eqref{eq:RatioTrick}, an estimator for our calculations of $S_2\left(A\right)$ is
\begin{equation}
    \label{eq:BoltzmannRT}
    S_2\left(A\right) = -\log \frac{Z\left[2, A \right]}{Z\left[2, \varnothing \right]}=-\log\left< e^{-\beta_c \left[E_{A}\left( \mathbf{s} \right) - E_{\varnothing}\left( \mathbf{s} \right)\right]}\right>_{\mathbf{s}\sim Z\left[2, \varnothing \right]},
\end{equation}
where $E$ is the Ising model energy and its subscript refers to the size of region $A$. A simplification is possible through the realization that the bulk contribution to the energy is exactly the same between $E_{A}$ and $E_{\varnothing}$. Therefore, the difference in energies of a configuration $\mathbf{s}$ is entirely due to the difference in energies of the inter-replica connections in region $A$. 

The exponential form of the estimator means that its value will be dominated by rare events. This problem is accentuated as the size of region $A$ increases leading to an exponential decrease in the performance. In order to combat such prohibitively inefficient scaling, one subdivides region $A$ into $N$ subregions $A_i$ such that $A_{i+1}$ contains $A_i$, and $A_{i+1} - A_{i} \leq \Delta A$, where $\Delta A$ is a hyper-parameter.  With these intermediate region $A$s, one can employ the so-called ``ratio'' or ``increment'' trick \cite{Hastings:2010} in order to express the desired ratio of partition functions as
\begin{equation}
    \label{eq:IncTrick}
    \frac{Z\left[2, A\right]}{Z\left[2, \varnothing \right]} = \prod_{i=0}^{N-1} \frac{Z\left[2, A_{i+1} \right]}{Z\left[2, A_{i}\right]},
\end{equation}
where $A_0 = \varnothing$ and $A_N= A$. Each term in this product is amenable to computation through the application of an estimator analogous to  Eq.~\eqref{eq:BoltzmannRT}, with $\mathbf{s}$ sampled from $Z\left[2, A_{i}\right]$ instead of $Z\left[2, \varnothing \right]$. With such a decomposition, we adjust the size of $\Delta A$ in order to control the variance of each estimator in the product, and we also give ourselves the option to parallelize the computation into $N$ separate processes.

We can express Eq.~\eqref{eq:IncTrick} in a conceptually useful way as 
\begin{equation}
    S_2(A)-S_2(A_0) = \sum_{i=0}^{N-1} \left[S_2(A_{i+1}) - S_2(A_{i}) \right],
\end{equation}
with $A_0=\varnothing$. However, this expression also generalizes to other choices of $A_0$. As each term in the sum is measured with a replica-based quantum Monte Carlo estimator analogous to Eq.~\eqref{eq:BoltzmannRT}, this expression reveals that the \Renyi entropy \textit{difference} between any pair of regions $A$ and $A_0$ can be estimated directly without the need to compute the values $S_2(A)$ and $S_2(A_0)$. This property can be exploited numerically in order to reduce the resulting uncertainty in the quantities of interest. Indeed, via a careful choice of $A_0$, contributions present in both $S(A)$ and $S(A_0)$ can be cancelled out exactly without affecting the statistical uncertainty of the desired quantities. In particular, for our calculations we choose $A_0$ such that the area law term is cancelled exactly and we directly probe the universal term $\chi_n$ (see Sec.~\ref{sec:Results}).
 
As a final means of improving the estimator for $S_2$, for the Ising model one can employ Fortuin-Kasteleyn cluster decomposition in order to derive a cluster version of the estimator in Eq.~(\ref{eq:BoltzmannRT}) such that\cite{Caraglio2008}
\begin{equation}
    \label{eq:LoopRT}
    \frac{Z\left[2, A_{i+1}\right]}{Z\left[2, A_{i} \right]} = \left<2^{n_{{}_{A_{i+1}}}\!\!\left(\mathbf{c}\right)\,-\,n_{{}_{A_{i}}}\!\!\left(\mathbf{c}\right)}\right>_{\mathbf{c}\sim Z\left[2, A_{i} \right]}.
\end{equation}
Here, $\mathbf{c}$ is a sampled configuration of Fortuin-Kasteleyn clusters and $n_{{}_{A_{i}}}\!(\mathbf{c})$ is the number of those clusters when the boundary conditions for region $A_{i}$ are imposed (with similar notation for $n_{{}_{A_{i+1}}}\!(\mathbf{c})$). We note that similar estimators can be derived in fully quantum simulations, such as within the framework of the Stochastic Series Expansion\cite{SSE, SSE2} for instance, as demonstrated in  Ref~\onlinecite{Bohdan}.

For Eq.~\eqref{eq:LoopRT}, a similar simplification occurs as that described for Eq.~\eqref{eq:BoltzmannRT}: the bulk clusters cancel out in the difference $n_{{}_{A_{i+1}}}\!(\mathbf{c})-n_{{}_{A_{i}}}\!(\mathbf{c})$, such that only the clusters that connect the replicas in region $\Delta A$ need to be built. One can achieve this simplification by starting the generation of every new cluster from spins located at the inter-replica connections and inside of  region $\Delta A$ until all of those spins are partitioned into corresponding clusters. Due to the non-local extent of the clusters on which it is built, this estimator has an exponentially improved performance with respect to the estimator based on Eq.~(\ref{eq:BoltzmannRT}). This improved performance increases the step size $\Delta A$ in Eq.~\eqref{eq:IncTrick} that can be used to achieve a targeted accuracy with given computational resources. Specifically, we find that we can push $\Delta A$ to values as large as $20$. This cluster-based estimator is the key ingredient that enables us to achieve the desired precision in the extraction of $\kappa_2$.

%% file: Results.tex
\section{Results}
\label{sec:Results}

We now turn to a discussion of our numerical extraction of $\kappa_2$. The raw data obtained from Monte Carlo simulations on the Ising model requires significant analysis due to a number of factors. In particular, the area law and the unknown conical singularity term pollute the universal contribution $\chi_2\!\left(u, b \right)$. Additionally, we are tasked with striking a fine balance when choosing the right cylinder height $\ell$. On one hand, we are interested in the thin-slice limit $\ell \ll L_x$, while on the other hand the continuum EE scaling is only expected to apply when $\ell \gg \delta$. For this reason, we first perform the scaling analysis on the free Gaussian scalar field theory on a lattice to benchmark against previously known results.\cite{Bueno:2015JHEP}  We then use this benchmark to aid in the analysis of the interacting Ising case. 

As discussed in Section \ref{Scaling}, we wish to perform fits to the function given in Eq.~\eqref{eq:Sn_scaling}. We consider toroidal rectangular geometries and fix $L_y = L$ and $L_x = 4L$, where the factor of 4 allows us to better probe the $\ell \ll L_x$ limit. In this thin-cylinder limit the universal geometric function is expected to take the form given in Eq.~\eqref{eq:Sn_thin} such that
\begin{equation}
\chi_2\!\left(u\rightarrow 0 \right) \sim -\kappa_2 \frac{L}{\ell},
\label{eq:f_thin}
\end{equation}
where we have removed the implicit dependence on the aspect ratio $b$ since we take it to be constant.

In order to remove the dominant area law contribution to the entropy scaling, we consider the difference in EE between two cylinders with different lengths $\ell$ and $\ell_0$, with $\ell < \ell_0$. In the case where the first cylinder is in the thin-slice regime such that $\ell \ll L_x$, we expect:
\begin{equation}
	S_2(\ell, L) - S_2(\ell_0, L) = -\kappa_2 \frac{L}{\ell} + g_2\!\left( \frac{\delta}{\ell} \right) - \underbrace{\chi_2\!\left( \frac{\ell_0}{L}  \right) - g_2\!\left( \frac{\delta}{\ell_0} \right)}_{\equiv B\left(\ell_0, L \right)}.
\label{eq:S2diff_scaling}
\end{equation}
As discussed in Sec.~\ref{subsec:EEEstimator}, this difference is directly measured in MC simulations, making our dataset completely free of the area law term and the associated statistical variance. We use the convention $\delta=1$ for our lattice calculations. 
\subsection{Free theory}\label{sec_free_thr}

\begin{figure}[t]
   \includegraphics{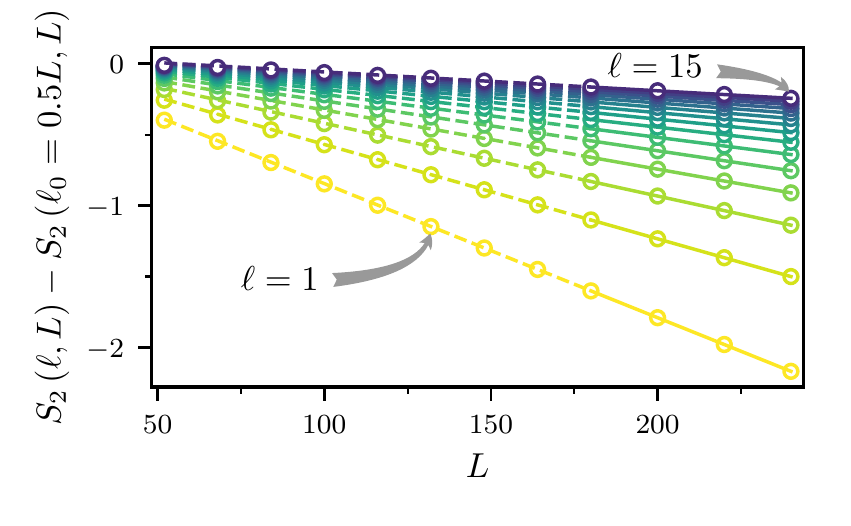}
\caption{Entanglement entropy with respect to a reference region for the free model as function of the system size. Different colours correspond to different sizes of region $A$ between $\ell = 1$ and $\ell = 15$. The solid lines are linear fits to four largest system sizes. The dashed continuations of the same color are the extrapolation of those fits to lower system sizes. The linear fit seems to capture most of dependence on $L$. Only upon closer examination can one observe that the extrapolation quality deteriorates for larger $\ell$, in line with the expectations discussed for the thin-slice limit in the main text.}
    \label{fig_free_data}
\end{figure}

We warm up with a free (Gaussian) scalar field theory on the toroidal system described above. In order to avoid zero modes that could lead to a logarithmic EE contribution, \cite{TwistTorus} we impose anti-periodic boundary conditions along the $y$-axis while periodic boundary conditions are kept for the $x$-direction.  The Hamiltonian is given by
\begin{equation}
H_{\text{free}} = 
\frac{1}{2}\sum_{i} \left( \pi_i^2 + m^2\phi_i^2 \right) 
+ \frac{1}{2} \sum_{\langle i, j \rangle} \left( \phi_i - \phi_j  \right)^2\, ,
\label{eq:Ham_free}
\end{equation}
where $\phi_i$ is a bosonic field with mass $m$ and conjugate momentum $\pi_i$.
Exact methods for calculating the \Renyi entropies for such free scalar theories are described in detail in Refs.~\onlinecite{Peschel, Casini:2009, HelmesPHD, LaurenPHD}, and have been employed previously by some of the authors in Refs.~\onlinecite{Helmes:2016,Witczak2017,Hayward2016,Hayward2017}. 

We plot in Fig.~\ref{fig_free_data} the left-hand side of Eq.~\eqref{eq:S2diff_scaling} for this free theory versus $L$, grouping points by their value of $\ell$. The behavior appears linear over a wide range of $\ell$ values, strongly hinting at a dominant contribution from the thin-cylinder form for $\chi_2(\ell/L)$ as in Eq.~\eqref{eq:f_thin}. 
Motivated by this observation, we perform fits of $S_2(\ell, L) - S_2(\ell_0, L)$ to a function $f_1$ that is linear in $\ell^{-1}$.
This is given by
\begin{equation}
    \label{eq:fit_lin}
    f_1\left(\ell \right) = -\kappa_2^{L}\frac{L}{\ell} + C_1,   
\end{equation}
for each system size $L$, where there are two fitting parameters $\kappa_2^{L}$ and $C_1$. The role of $C_1$ is to absorb the offset due to the $\ell$-independent term  $B\left(\ell_0, L \right)$. When performing these and the following fits, instead of fitting all points at once, we fit data over a sliding window. For this reason, the estimate for $\kappa_2$ has an explicit dependence on both $\ell$ and $L$. However, to reduce the notational clutter, we avoid showing the dependence on $\ell$, since the $x$-axis in all the figures makes this dependence explicit. 

\begin{figure}[t]
   \includegraphics{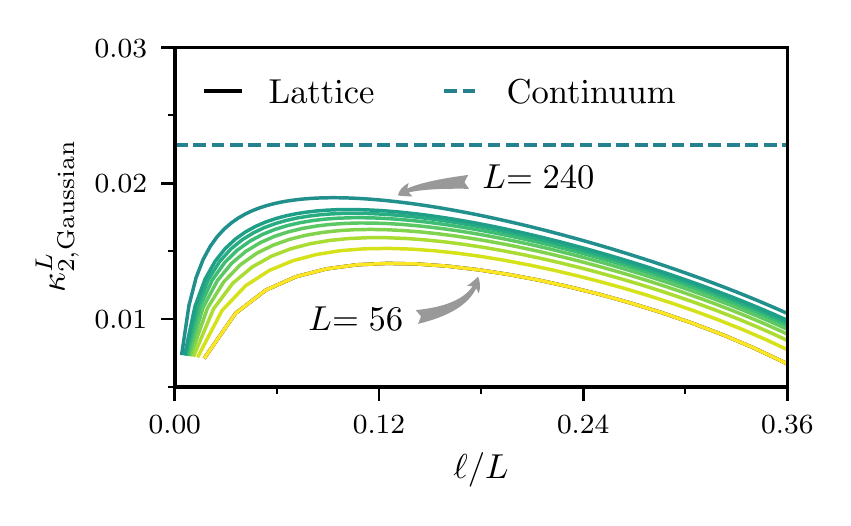}
\caption{$\kappa_2$ values for the free theory extracted from Fig. \ref{fig_free_data} via the linear fit in Eq.~\eqref{eq:fit_lin}, for different system sizes $L$. The $x$-axis indicates the lowest value of $\ell$ used in each six-point fit divided by the system size. The density of the extracted values on the $x$-axis is so high that plotting each point results in too much clutter and thus a continuous line representation is used instead. The dashed line represents the theoretical value for the Gaussian theory in the thermodynamic limit.\cite{Bueno:2015JHEP}}
    \label{fig_free_linear}
\end{figure}

The results for the extracted $\kappa_{2,\text{Gaussian}}^L$ are shown in Fig.~\ref{fig_free_linear}.  This plot reveals non-linear dependencies on $\ell^{-1}$ and demonstrates the challenge in extracting an unbiased estimate for $\kappa_2$. 
Most notably, as judged by the proximity to the known exact result from the Gaussian theory, the best estimate for $\kappa_2^L$ does not come from the thinnest cylinders with $\ell=\delta=1$. This observation is not surprising since the EE scaling prediction is only expected to hold in the continuum, which, on a lattice, amounts to the requirement $\delta \ll \ell$. Both this condition and the thin-slice requirement $\ell \ll L$ constitute the challenge of tuning to a regime where both criteria are satisfied. One possible approach is to consider the largest extracted $\kappa_2^L$ as the best estimate for each system size. However, these values converge quickly to an underestimate for the true value, implying that this extrapolation technique will yield a biased result.

An alternative approach originates from the following insight: any length on a lattice is only defined up to the lattice spacing. As a consequence, on a lattice it is impossible to distinguish between the class of continuum cylindrical regions $A$ with lengths in the range $(\ell-\delta, \ell+\delta)$. In light of this realization, let us 
introduce a parameter $\gamma \in \left(-\delta, \delta \right)$ designed to capture the degree of freedom associated with mapping the lattice and continuum theories such that
the left-hand side of Eq.~\eqref{eq:S2diff_scaling} instead takes the form $-\kappa_2 \frac{L}{\ell+\gamma}  + g_2\left( \frac{\delta}{\ell} \right) + B\left(\ell_0, L \right)$.  Here we have included $\gamma$ only within the term for which it contributes most significantly.
Notice that with $L$ and $\ell_0$ kept constant, $B\left(\ell_0, L \right)$ is also a constant and, therefore, can be ignored in the following discussion. The expected first-order offset due to $\gamma$ is then $-\frac{L}{\ell}\left(\kappa_2- \kappa_2\frac{\gamma}{\ell} - \frac{g_2\left( \delta/\ell\right)\ell}{L} \right)$, 
and the expression in the parentheses is a good approximation for the value of $\kappa_2^L$ extracted under a linear fit to Eq.~\eqref{eq:fit_lin}. This expression thus explains well the biases observed in Fig.~\ref{fig_free_linear}. Namely, we observe that the estimate  for $\kappa_2$ is polluted by two terms. While the term containing the conical singularity scaling function $g_2\left( \delta/\ell\right)$ can be eliminated by taking the thermodynamic limit $L \to \infty$, the systematic offset due to $\gamma$ can only be suppressed by scaling $\ell$ to infinity as well. In addition, this term explains the observed increased negative offset in the regime $\ell\sim\delta$. 

\begin{figure}[t]
   \includegraphics{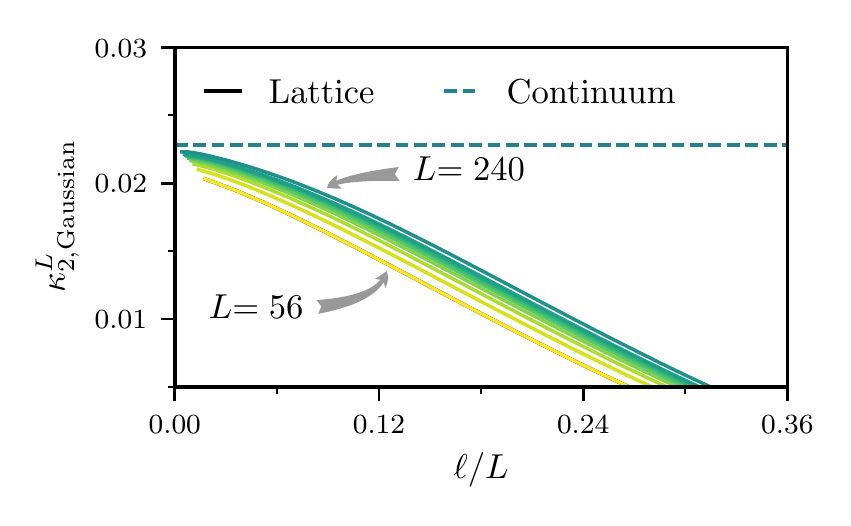}
\caption{$\kappa_2$ values for the free theory extracted from Fig. \ref{fig_free_data} via the non-linear fit in Eq.~\eqref{eq:fit_inv}, with an additional parameter $\gamma$ introduced to capture the short-distance behavior important in the regime of small $\ell$. Each fit again uses six points, and the value of $\ell$ for the $x$-axis corresponds to the lowest value used in each fit. The dashed line indicates the known value for the Gaussian theory in the thermodynamic limit.\cite{Bueno:2015JHEP}}
    \label{fig_free_inverse}
\end{figure}

In order to take these considerations into account, we parameterize the ambiguity in the definition of the cylinder's length by including the parameter $\gamma$ in out fits. We perform new non-linear fits of $S_2(\ell, L) - S_2(\ell_0, L)$ to the function
\begin{equation}
    \label{eq:fit_inv}
    f_2\left(\ell\right) = -\kappa_2^{L}\frac{L}{\ell+\gamma} + C_2,
\end{equation}
for each $L$, now with three fitting parameters $\kappa_2^{L}$, $\gamma$ and $C_2$. The outcome for the free theory is illustrated in Fig.~\ref{fig_free_inverse}. The dependence of the extracted $\kappa_2^{L}$ estimate on $\ell$ for this theory has now drastically changed. In particular, the introduction of the parameter $\gamma$ has completely removed the downward drop observed at small $\ell$ in Fig.~\ref{fig_free_linear}. Moreover, the systematic offset, \textit{i.e.},~the difference between the peak and the known value,  which seemed to survive to the thermodynamic limit in Fig.~\ref{fig_free_linear}, is also remedied.

\begin{figure}[t]
   \includegraphics{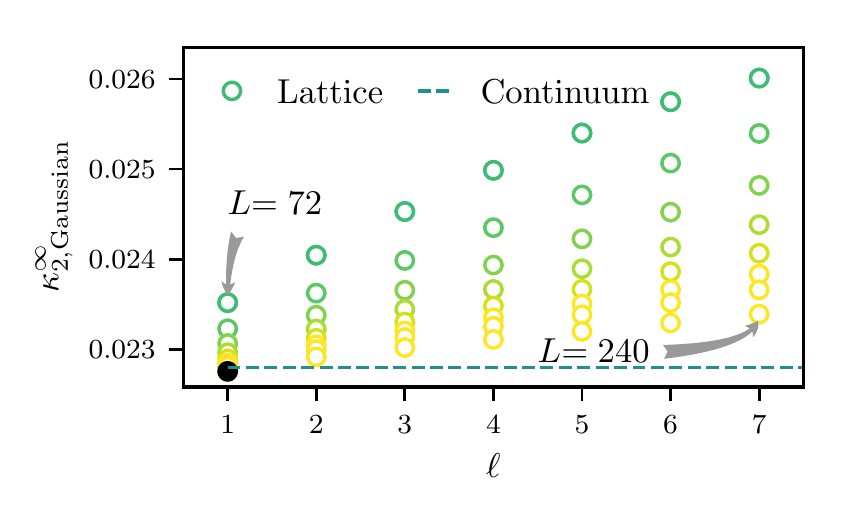}
\caption{$\kappa_2$ values for the free theory extrapolated to the thermodynamic limit. These values are obtained via an additional six-point fit to the form in Eq.~\ref{eq:fit_ext} to the results in the Fig. \ref{fig_free_inverse}. The $x$-axis indicates the value of $\ell$ used in each fit. The dashed line represents the known continuum value for the Gaussian theory\cite{Bueno:2015JHEP}, while the solid black circle indicates our best numerical estimate, namely $\kappa_{2,\text{Gaussian}}^{\infty} = 0.0227558$, for the free theory in the thermodynamic limit.}
    \label{fig_free_extrap}
\end{figure}

With a firmly-supported understanding of the short-distance scaling of the \Renyi entropy, we are ready to account for the conical singularity in our estimate for $\kappa_2$. For this we note that although the non-linear fit based on Eq.~\eqref{eq:fit_inv} cannot distinguish the contributions from the universal term and the conical singularity, the latter is independent of system size and therefore its relative magnitude decays as $L^{-1}$. To be more specific,  $\kappa_2^{L} \approx \kappa_2- \frac{g_2\left( \delta/\ell\right)\ell}{ L} $.
Consequently, we perform a second fit, extrapolating the previously extracted $\kappa_2^{L}$ towards $L=\infty$ via a two-parameter fit linear in $L^{-1}$ such that, for each $\ell$, we fit the results from Fig.~\ref{fig_free_inverse} to the function
\begin{equation}
    \label{eq:fit_ext}
    f_{\text{extrap.}}\left(L\right) = -C_{\text{extrap.}}/L + \kappa_2^{\infty},
\end{equation}
where $\kappa_2^{\infty}$ and $C_{\text{extrap.}}$ are fitting parameters. Here $\kappa_2^{\infty}$ represents our final estimate for $\kappa_2$ extrapolated to the thermodynamic limit, with the results for the free theory illustrated in Fig. \ref{fig_free_extrap}. Taking the value corresponding to the smallest region $A$ on the largest system considered as our best numerical estimate, we find $\kappa_{2,\text{Gaussian}}^{\infty} = 0.0227558$, which is less than $0.2\%$ below the known  value  of 0.0227998.\cite{Bueno:2015JHEP} To put this result into context, we can compare our estimate to that obtained in Ref.~\onlinecite{Hayward2016}, which also aims to numerically extract $\kappa_{2,\text{Gaussian}}$ but does not take into account the proposed offset that scales as $\frac{\gamma}{\ell}$. Although the authors were able to collect a dataset for system sizes as large as $L/\delta=2000$, their extrapolation to the thermodynamic limit yielded a value for $\kappa_{2,\text{Gaussian}}$ that is still $9\%$ off from its theoretical value. Therefore, our fitting procedure produces an estimate which is two orders of magnitude closer to the theoretical value, despite considering system sizes an order of magnitude smaller.

\subsection{Interacting theory}

\begin{figure}[t]
   \includegraphics{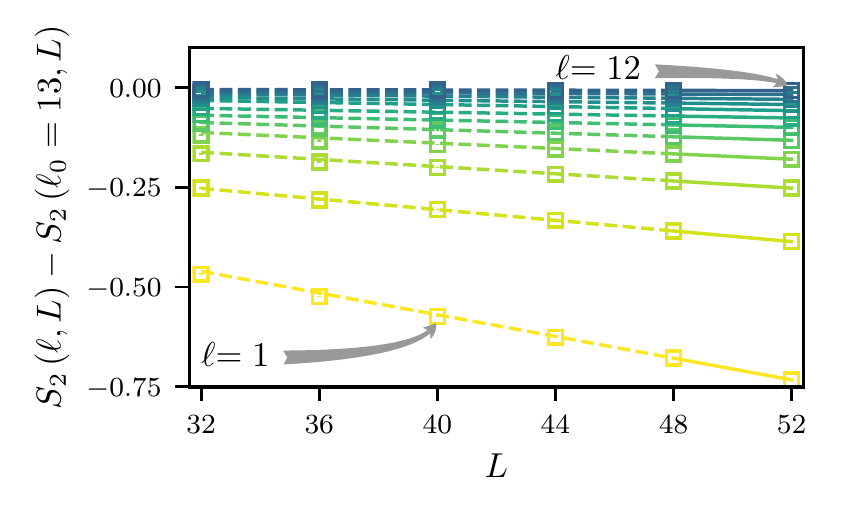}
\caption{Entanglement entropy with respect to a reference region $A$ of size $\ell_0=13$ for the 2d Transverse Field Ising model at criticality as a function of the system size. Different colours correspond to different sizes of region $A$ from $\ell = 1$ to $\ell = 12$. The solid lines are linear fits to the two largest system sizes, and the dashed continuations of the same colour are the extrapolations of those fits to lower system sizes. As in Fig.~\ref{fig_free_data}, the linear fit seems to capture most of the dependence on $L$, and it is only upon closer examination that one can appreciate the need for the two-step fitting procedure developed in Sec.~\ref{sec_free_thr}. }
    \label{fig_inter_data}
\end{figure}

We now use the insight gained in the previous section to proceed with the extraction of $\kappa_{2,\text{WF}}$ for the interacting Ising theory at the Wilson-Fisher fixed point. As before, we begin by plotting the full data set as a function of $L$ (see Fig. \ref{fig_inter_data}), and we again note a dominant linear behaviour attributed to the universal term's contribution in the thin-slice limit. This observation suggests that the two-step fitting procedure developed above for the free theory can be carried over and applied for this interacting data set. 

\begin{figure}[t]
   \includegraphics{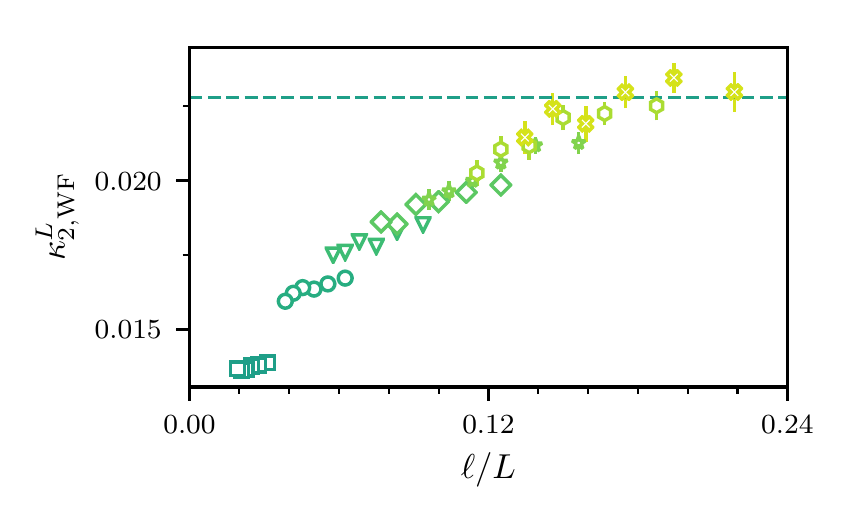}
\caption{$\kappa_2$ values for the Ising theory extracted from Fig.~\ref{fig_inter_data} via the linear fit in Eq.~\ref{eq:fit_lin}, for different system sizes $L$. Markers with the same color and style share the same value of $\ell$. The $x$-axis indicates the lowest value of $\ell$ used in each six-point fit divided by the system size. The dashed line represents the theoretical value for the Gaussian theory in the thermodynamic limit.\cite{Bueno:2015JHEP}}
    \label{fig_inter_linear}
\end{figure}

\begin{figure}[t]
   \includegraphics{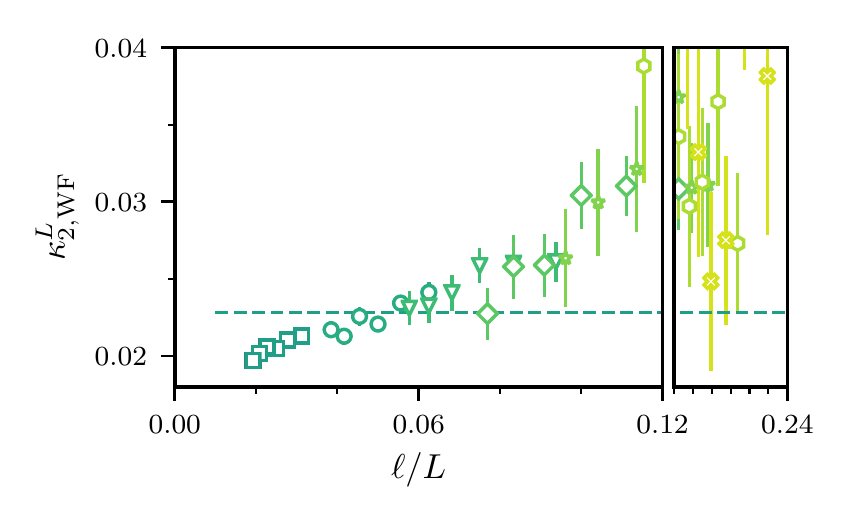}
\caption{$\kappa_2$ values for the Ising theory extracted from Fig.~\ref{fig_inter_data} via the non-linear fit in Eq.~\ref{eq:fit_inv}, with each fit using six point. Here we include the additional parameter $\gamma$ introduced to capture the short-distance behavior important in the regime of small $\ell$. Markers with the same color and style share the same value of $\ell$. The value of $\ell$ for the $x$-axis indicates the lowest value of $\ell$ used in each fit divided. The dashed line represents the theoretical value for the Gaussian theory in the thermodynamic limit.\cite{Bueno:2015JHEP} The $x$-axis is split into two parts in order to focus on the values of the extracted $\kappa_{2,\text{WF}}^L$ at low $\ell/L$, since the values are noisy for $\ell/L>0.12$.} 
    \label{fig_inter_inverse}
\end{figure}

The effect of including the fit parameter $\gamma$ for extracting an unbiased estimate for $\kappa_{2,\text{WF}}$ can be seen by comparing fits with (Fig.~\ref{fig_inter_linear}) and without it (Fig. \ref{fig_inter_inverse}). The estimates obtained via fits containing $\gamma$ are all above the corresponding estimates without it. This situation is analogous to the systematic bias towards lower values of the extracted $\kappa_{2,\text{Gaussian}}$ observed for the free theory when the $\gamma$ parameter is not included in the fits (see Fig.~\ref{fig_free_linear} and Fig.~\ref{fig_free_inverse}). 

\begin{figure}[t]
   \includegraphics{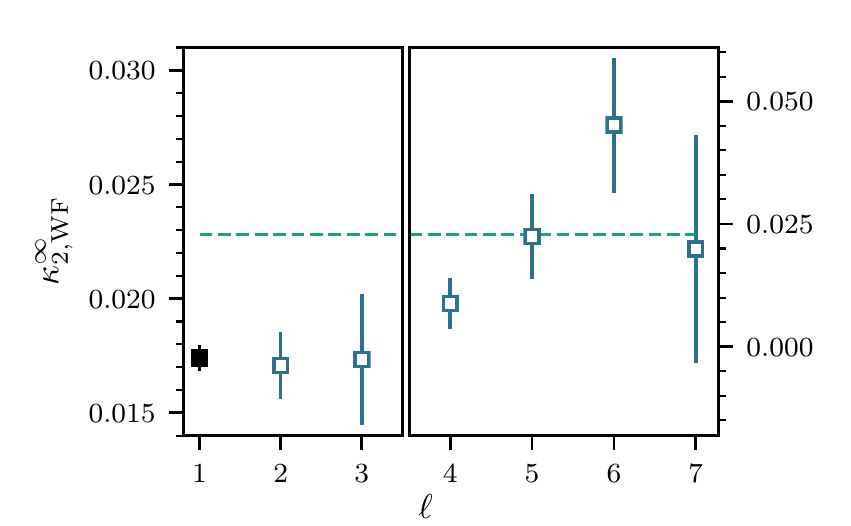}
\caption{$\kappa_2$ values for the Ising theory extrapolated to the thermodynamic limit. These values are obtained by fitting the results from Fig.~\ref{fig_inter_inverse} to form in Eq.~\ref{eq:fit_ext}. The $x$-axis indicates the $\ell$ value used in each six-point fit. The black marker for $\ell=1$ indicates our best numerical estimate $\kappa_{2,\text{WF}}^{\infty} =0.0174(5)$, and the dashed line represents the known value for the Gaussian theory in the thermodynamic limit.\cite{Bueno:2015JHEP} The plot is split into two halves, with the right half providing a wider range of $y$ values. These values are not as accurate as the ones on the left half but show the trend of decreasing error-bars and a decreasing estimate for $\kappa_{2,\text{WF}}^\infty$.}
    \label{fig_inter_extrap}
\end{figure}

Concentrating further on the fits including the parameter $\gamma$ in Fig.~\ref{fig_inter_inverse}, we note a systematic increase in error bar with increasing  $\ell/L$. This trend can be explained by considering the relative strength of the leading contribution to the universal term (proportional to $L/\ell$), against the subleading terms. These terms originate from contributions due to the conical singularity and the next-order Taylor expansion in $\chi_2\left( u\right)$, and scale like $u=\ell/L$ relative to the leading term. Since our fitting form ignores these additional terms, the error bars can be seen as a qualitative indicator for the validity of 
the assumption of their relative insignificance.
Indeed, the regime where the error bars are empirically small is also that in which we are interested ($\ell/\delta \rightarrow 1$), allowing us to proceed with confidence onto the second extrapolation that estimates $\kappa_{2,\text{WF}}$. This step is completely analogous to that done in the free theory extrapolation (see Sec. \ref{sec_free_thr}), and our results for the Ising theory are shown in Fig.~\ref{fig_inter_extrap}. As before, this step is based on the fact that the relative strength of the previously neglected terms decay as $\delta/L$. 
In correspondence with our previous discussion, the error bars are significantly reduced for small $\ell$. Furthermore, we note that for a decreasing $\ell$ the value for $\kappa_{2,\text{WF}}^{\infty}$ seems to decrease at first, similar to the trend observed for the free theory in Fig.~\ref{fig_free_extrap}. However, our $\kappa_{2,\text{WF}}^{\infty}$ estimates seem to stabilize within error-bars for $\ell \leq 3$. This fact can be interpreted as an indication that we have reached system sizes large enough to accurately probe the $u\rightarrow 0$ limit. Concluding our analysis, we take the extrapolated value $\kappa_{2,\text{WF}}^{\infty} = 0.0174(5)$, which is extracted at $\ell=1$, as our best estimate for $\kappa_2$ at the Wilson-Fischer fixed point.

%% file: Discussion.tex
\section{Discussion}
\label{sec:Discussion}
We have studied 
a novel universal quantity, $\kappa_2$, which quantifies a divergence of the \Renyi entanglement entropy when a torus is bipartitioned into two cylinders.  The divergence appears subleading to the area law, in the limit when one of the cylinder lengths approaches zero.
We have performed a Monte Carlo calculation on a $2+1$ dimensional Ising model to discover the value of $\kappa_{2,\text{WF}}$ for the interacting Wilson-Fisher fixed point.

Our calculations are performed on the lattice and require careful control of finite-size effects in the scaling of the \Renyi entropy.  In order to benchmark our novel fitting procedure used to extract the subleading divergence, we first performed calculations for a free scalar field theory, regularized on finite-size lattices.  In this case, the thermodynamic value of $\kappa_{2,\text{Gaussian}}$ was calculated previously in the continuum, which allows us to test our fitting procedures based against a known exact result. This benchmark illuminates the crucial importance of the fitting parameter $\gamma$ that we introduce in order to extend the range of applicability of continuum results to the lattice. 
Surprisingly, this parameter enables us to take advantage of more of our dataset, increasing the efficiency of our calculations in thin-slice regime. 
For the second \Renyi entropy of the free field, we extract the value $\kappa_{2,\text{Gaussian}} = 0.0227558$, which is accurate to within $0.2 \%$ of the continuum value.

We then performed extensive Monte Carlo simulations on the interacting Ising model, using the same lattice geometries as in the analysis of the free theory.
The use of an improved \Renyi entropy estimator based on Fortuin-Kastelyn cluster decomposition in conjunction with the ratio trick proves to be fundamental in obtaining the accuracy required for a reliable analysis.\cite{Bohdan}
After the extrapolation to the thermodynamic limit, our best estimate for the value of this universal coefficient at the interacting (Wilson-Fisher) fixed point is $\kappa_{2,\text{WF}}=0.0174(5)$. In total, extracting this value used approximately 200 years of CPU time.

The coefficient $\kappa_2$ that we have extracted from cylindrical geometries in the thin-slice limit additionally serves to give insight about entangling geometries with corners that are difficult to access through direct means on a lattice.
Specifically, $\kappa_2$ is related to the logarithmic coefficient $a_2(\theta)$ that arises in the scaling of the \Renyi EE when the entangling geometry contains a corner such that, in the small-angle limit, $a_2(\theta \to 0) = \kappa_2 / \theta$.\cite{Casini:2009, CornerCylinder}
Further, combining our results for $\kappa_{2,\text{WF}}$ with previous results for $a_{2,\text{WF}}(\pi/2)$ in the same interacting theory, we expect that one can approximately reconstruct the behaviour of $a_{2,\text{WF}}(\theta)$ for all angles $\theta$ using techniques similar to those proposed in Ref.~\onlinecite{Bueno:2015JHEP}. 

Our numerical value of $\kappa_2$ for the interacting fixed point is relatively close to the value for the free theory, but does have a  significant difference when statistical errors are taken into account.  
This finding is interesting in the context of a recent large-$N$ calculation for the more general $N$-component $O(N)$ model,\cite{wFlargeN} 
in which the $\kappa_1$ value, extracted from the von Neumann EE, of the Wilson-Fisher fixed point is expected to be directly related to the non-interacting Gaussian fixed point as $\kappa_{1,\text{WF}}(N) \simeq N\kappa_{1,\text{Gaussian}}$ to leading order in $N^{-1}$.
This theoretical prediction is only applicable to $\kappa_1$, thus leaving an intriguing possibility that $\kappa_2$ captures non-trivial differences between the Gaussian and Wilson-Fisher fixed points that extend to the large-$N$ limit.
In addition, the observed proximity of $\kappa_{2,\text{WF}}$ to $\kappa_{2,\text{Gaussian}}$ for the Ising ($N=1$) case is similar to the situation encountered for the universal coefficient of the logarithmic scaling term that arises due to a corner in an entangling boundary in $2+1$.  There, the interacting value is numerically very close to the free Gaussian theory.  This behaviour changes for $N>1$, where extensive calculations show the universal coefficient increasing with $N$.
An interesting avenue for future work would be to examine if $\kappa_2$ obeys a similar trend.

The effort required by our numerical strategy illustrates the challenge implicit in the calculation of universal quantities in the entanglement entropy for interacting theories in $2+1$ dimensions.  
Indeed, a high degree of numerical precision is required if such universal numbers are to be of potential use in future studies of novel interacting lattice models, where they may be useful in identifying underlying field theories and their emergent degrees of freedom,\cite{Kallin:2014} and possibly even in attaching an organizational hierarchy to the associated fixed points.\cite{Grover:2014e}
The program to calculate such universal quantities in a wide variety of interacting critical theories will continue with concentrated numerical studies for many years to come.